# Competing edge structures of Sb and Bi bilayers by trivial and nontrivial band topologies

Sang-Min Jeong[1], Seho Yi[1], Hyun-Jung Kim,[2,1] Gustav Bihlmayer,[3] and Jun-Hyung Cho[1*]

[1] *Department of Physics, Research Institute for Natural Science,*
*and HYU-HPSTAR-CIS High Pressure Research Center,*
*Hanyang University, 222 Wangsimni-ro, Seongdong-Ku, Seoul 04763, Korea*
[2] *Korea Institute for Advanced Study, 85 Hoegiro, Dongdaemun-gu, Seoul 02455, Korea*
[3] *Peter Grünberg Institut and Institute for Advanced Simulation,*
*Forschungszentrum Jülich and JARA, 52425 Jülich, Germany*
(Dated: June 12, 2018)

One-dimensional (1D) edge states formed at the boundaries of 2D normal and topological insulators have shown intriguing quantum phases such as charge density wave and quantum spin Hall effect. Based on first-principles density-functional theory calculations including spin-orbit coupling (SOC), we show that the edge states of zigzag Sb(111) and Bi(111) nanoribbons drastically change the stability of their edge structures. For zigzag Sb(111) nanoribbon, the Peierls-distorted or reconstructed edge structure is stabilized by a band-gap opening. However, for zigzag Bi(111) nanoribbon, such two insulating structures are destabilized due to the presence of topologically protected gapless edge states, resulting in the stabilization of a metallic, shear-distorted edge structure. We also show that the edge states of the Bi(111) nanoribbon exhibit a larger Rashba-type spin splitting at the boundary of Brillouin zone, compared to those of the Sb(111) nanoribbon. Interestingly, the spin textures of edge states in the Peierls-distorted Sb edge structure and the shear-distorted Bi edge structure have all three spin components perpendicular and parallel to the edges, due to their broken mirror-plane symmetry. The present findings demonstrate that the topologically trivial and nontrivial edge states play crucial roles in determining the edge structures of normal and topological insulators.

PACS numbers:

## I. INTRODUCTION

Over the last decade, a number of two-dimensional (2D) materials have been discovered because of their intriguing emergent electronic properties that can be exploited for various device applications[1–5]. Specifically, one-dimensional (1D) edge states formed at the boundaries of some 2D materials showed the delicate coupling and competitions of charge, spin, and lattice degrees of freedom, thereby giving rise to interesting quantum phases such as charge- or spin-density waves and magnetism[6–12]. Contrasting with these macroscopic quantum condensates characterized by spontaneous symmetry breaking, the quantum spin Hall (QSH) effect in 2D topological insulators provides a different paradigm in classifying edge states based on the topological orders[13,14]. The edge states of 2D topological insulators feature helical gapless boundary states that are protected by time-reversal symmetry[15]. Thus, edge states in 2D normal and topological insulators can offer unique playgrounds for exploration of novel, distinctive quantum phases.

To demonstrate the drastically different features of edge states in 2D materials, we here consider the Sb(111)[16,17] and Bi(111)[18–26] bilayers that have been known to be the normal and topological insulators, respectively. The zigzag nanoribbons of the two bilayers have a dangling-bond (DB) electron per each edge atom, giving rise to the formation of 1D edge states that are half-filled and cross the Fermi level $E_F$ at $k_F = \pi/2a_x$ ($a_x$ is the lattice parameter along the direction parallel to the edges). For zigzag Sb(111) nanoribbon (designated as ZSNR), such half-filled bands may induce an electronic instability of charge density wave (CDW)[6,7]. Here, the CDW with the wave vector of $2k_F$ tends to change the periodicity of edge structure through electron-lattice coupling, thereby opening a band gap at the Brillouin-zone boundary of a new doubled unit cell. Meanwhile, for zigzag Bi(111) nanoribbon (ZBNR), such a metal-insulator transition due to the Fermi surface nesting should be suppressed by the QSH effect that has topologically protected gapless edge states in the bulk band gap. Consequently, the contrasting features of the edge states of ZSNR and ZBNR are expected to influence their edge structures in the different ways.

In the present study, we demonstrate the stability of different edge structures in the Sb(111) and Bi(111) bilayers using density-functional theory (DFT) calculations with the inclusion of spin-orbit coupling (SOC). For ZSNR, we find that the edge states are susceptible to a Fermi surface nesting-induced CDW formation, leading to a Peierls-distorted edge (PE) structure with a gap opening. This CDW formation is found to be further stabilized by a lattice reconstruction converting two edge hexagons into a pentagon-heptagon pair, which enlarges the band-gap opening. By contrast, ZBNR is found not to favor a Peierls distortion or reconstruction because of the presence of topologically protected gapless edge states, but to stabilize a metallic shear-distorted edge (SE) structure where edge atoms are laterally displaced parallel to the edges. Therefore, the edge structures of ZSNR and ZBNR are drastically changed by their topologically trivial and nontrivial edge states, respectively. Interestingly, the PE or SE structure with broken mirror-plane symmetry exhibits an intricate spin texture of edge states containing all three spin components perpendicular and parallel to the edges, which can be associated with the unquenched orbital angular momentum at the edge.

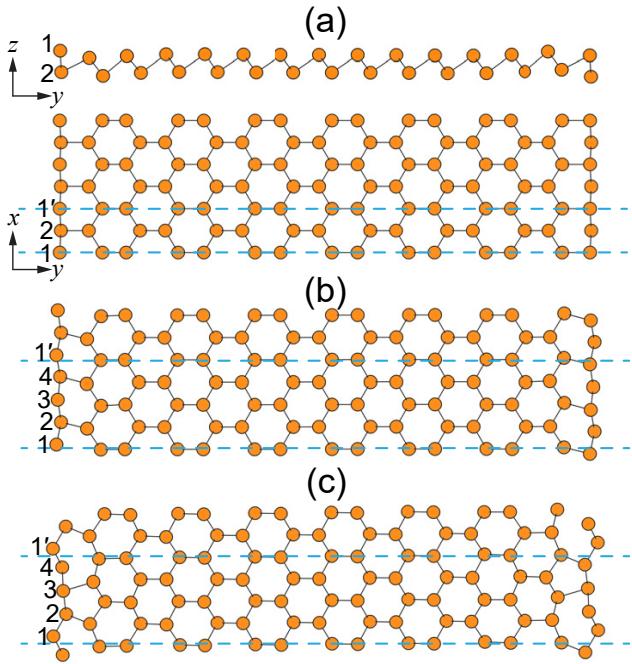

FIG. 1: (Color online) (a) Side and top views of the optimized structure of ZSNR-BE. The top views of ZSNR-PE and ZSNR-RE are given in (b) and (c), respectively. The dashed lines indicate the unit cell of each structure. The edge Sb atoms at one edge are labeled.

## II. COMPUTATIONAL METHODS

The present DFT calculations were performed using the Vienna *ab initio* simulation package with the projector-augmented wave method[27–29]. For the exchange-correlation energy, we employed the generalized-gradient approximation functional of Perdew-Burke-Ernzerhof (PBE)[30]. The edges of the Sb(111) and Bi(111) bilayers were modeled by a periodic supercell consisting of zigzag nanoribbons, each of which has 30 Sb and 30 Bi atoms within the 1×1 unit cell, respectively. Here, neighboring nanoribbons are separated by ∼20 Å of vacuum along the $y$ and $z$ directions [see Fig. 1(a)], making the interribbon interactions negligible. We simulated the bulk-truncated edge (BE) or SE structure using the 1×1 unit cell, while the PE and reconstructed edge (RE) structures using the 2×1 unit cell: see Figs. 1(a), 1(b), and 1(c), respectively. It is noted that the BE structure has mirror symmetry through the $yz$ plane, while the other edge structures break the mirror symmetry. A plane-wave basis was employed with a kinetic energy cutoff of 400 eV, and the **k**-space integration was done with 48 and 24 points in the Brillouin zone of the 1×1 and 2×1 unit cells, respectively. All atoms were allowed to relax along the calculated forces until all the residual force components were less than 0.005 eV/Å. For the Sb(111) and Bi(111) bilayers, we obtain the optimized lattice parameters $a_x$ = 4.114 (4.121) and 4.336 (4.389) Å using the PBE (PBE+SOC) calculation, respectively, in good agreements with previous DFT calculations[16,17,23,25].

## III. RESULTS

### 1. Zigzag Sb nanoribbon

We first optimize the BE structure of ZSNR (hereafter termed ZSNR-BE) using the PBE calculation without the inclusion of SOC. Figure 1(a) shows the side and top views of the optimized structure of ZSNR-BE. We find that the outermost edge Sb atoms [$Sb_1$ in Fig. 1(a)] relax inward significantly with displacement along the $y$ and $z$ directions. Consequently, the bond length $d_{Sb_1-Sb_2}$ of edge atoms is contracted to be 2.856 Å (see Table I), shorter than that (2.890 Å) of bulk Sb atoms. The band structure of ZSNR-BE is given in Fig. 2(a), showing the presence of half-filled bands with a bandwidth of ∼0.78 eV. Obviously, the band projection onto the 5$p$ orbitals of $Sb_1$ and $Sb_2$ atoms demonstrates that the half-filled bands originate mostly from the edge atoms [see Fig. 2(a)]. We note that the two edge states arising from the opposite edges cross $E_F$ at the almost midpoints of the positive and negative ΓX symmetry lines, indicating a Fermi-surface-nesting vector of $2k_F = \pi/a_x$. As shown in Fig. 3(a), the charge character of edge state at the X point represents a large localization of DB electrons around the $Sb_1$ atoms with some penetration into the center of the nanoribbon. Such a penetration is found to be more enhanced at the Γ point [see Fig. 3(b)], giving rise to an interedge interaction. Consequently, the two edge states are hybridized to open a small gap of 14 meV at the Γ point [see the inset of Fig. 2(a)].

TABLE I: Bond lengths (in Å) of edge atoms in various structures of ZSNR and ZBNR, obtained without and with the inclusion of SOC. The labels of edge atoms are shown in Fig. 1.

|  |  | $d_{1-2}$ | $d_{2-3}$ | $d_{3-4}$ | $d_{4-1'}$ |
|---|---|---|---|---|---|
| w/o SOC | ZSNR-BE | 2.856 | – | – | – |
|  | -PE | 2.874 | 2.815 | 2.879 | 2.851 |
|  | -RE | 2.886 | 2.914 | 2.865 | 2.712 |
| w SOC | ZSNR-BE | 2.859 | – | – | – |
|  | -PE | 2.877 | 2.818 | 2.882 | 2.853 |
|  | -RE | 2.890 | 2.918 | 2.868 | 2.711 |
| w/o SOC | ZBNR-BE | 3.006 | – | – | – |
|  | -PE | 3.035 | 3.009 | 3.026 | 2.964 |
|  | -RE | 3.049 | 3.055 | 3.022 | 2.879 |
| w SOC | ZBNR-BE | 3.033 | – | – | – |
|  | -SE | 3.049 | 3.020 | – | – |
|  | -RE | 3.072 | 3.074 | 3.045 | 2.912 |

Due to the Fermi surface nesting in ZSNR-BE, electrons and holes near $E_F$ can couple with a lattice vibration of wave length $2a_x$, thereby giving rise to the so-called Peierls lattice distortion as well as a band-gap opening at the new Brillouin zone boundary[6,7]. Consequently, the PE structure of ZSNR (ZSNR-PE) is found to be more stable than ZSNR-BE by 2.5 meV/atom [see Fig. 4(a)]. The Peierls lattice distortion is represented by a bond-length alternation with $d_{Sb_1-Sb_2}$ = 2.874

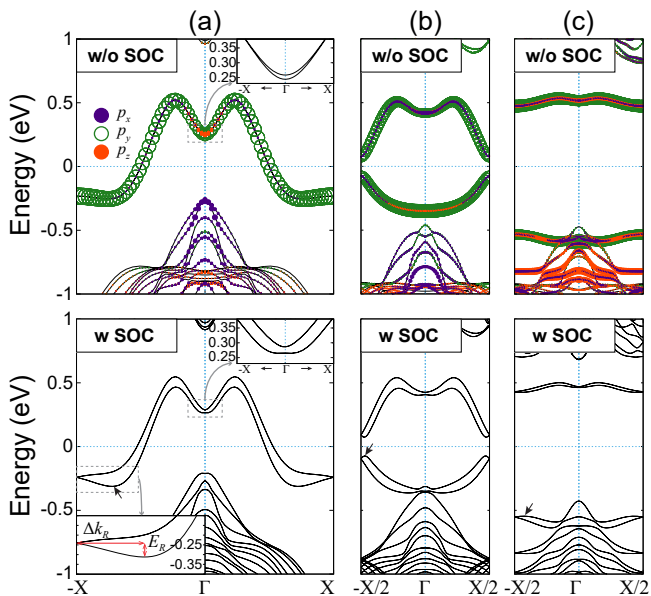

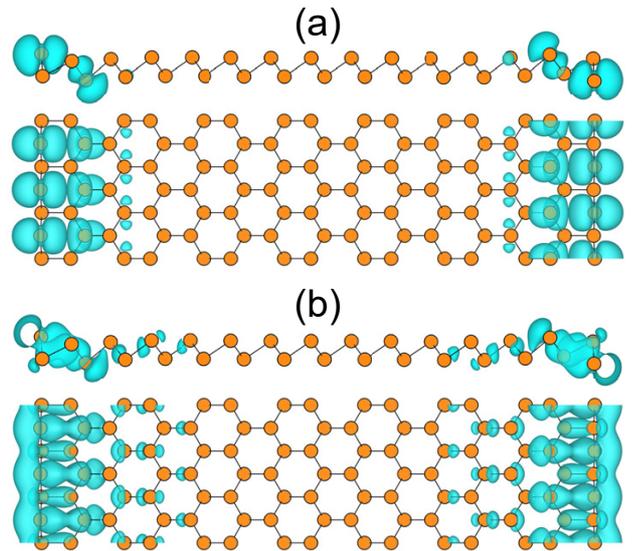

FIG. 2: (Color online) Band structures of (a) ZSNR-BE, (b) ZSNR-PE, and (c) ZSNR-RE, obtained without and with the inclusion of SOC. The energy zero represents the Fermi level. The direction of the $\Gamma-X$ line is parallel to the edges. The insets in (a) magnify the band dispersions near the $\Gamma$ point. The circles represent the bands projected onto the $5p_x$, $5p_y$ and $5p_z$ orbitals of the edge $Sb_1$ and $Sb_2$ atoms. Here, the circle radii are proportional to the weights of the corresponding orbitals. The arrows in the lower panels indicate the position of energy minimum or maximum around which the Rashba parameter $\alpha_R$ is evaluated.

FIG. 3: (Color online) Side and top views of the charge densities of the ZSNR-BE edge state at the (a) X and (b) $\Gamma$ points. The charge densities are drawn with an isosurface of $5\times10^{-4}$ electrons/Å.

Å, $d_{Sb_2-Sb_3}$ = 2.815 Å, $d_{Sb_3-Sb_4}$ = 2.879 Å, and $d_{Sb_4-Sb_{1'}}$ = 2.853 Å [see Fig. 1(b) and Table I]. Figure 2(b) shows the calculated band structure of ZSNR-PE, where the edge states open a band gap of 149 meV at the X point. It is notable that the zigzag graphene nanoribbon has 1D edge states which stabilize ferromagnetic spin ordering at each edge[31,32]. In order to examine the stability of magnetic order in ZSNR-BE, we perform the spin-polarized calculations for various ferromagnetic and antiferromagnetic configurations within the 1×1 and 2×1 unit cells, respectively. However, we were not able to find any spin ordering in ZSNR-BE.

We further investigate another edge structure of ZSNR by taking into account a reconstruction: i.e., two hexagons in the edges are converted into a pentagon-heptagon pair, which is similar to the structure of the Stone-Wales defect in graphene nanoribbons[33,34]. The optimized RE structure of ZSNR (ZSNR-RE) is displayed in Fig. 1(c). We find that ZSNR-RE is more stable than ZSNR-BE (ZSNR-PE) by 4.6 (2.1) meV/atom [see Fig. 4(a)]. This enhanced stability of ZSNR-RE is likely caused by the saturation of DBs at the edges, forming a double bond between the $Sb_4$ and $Sb'_1$ atoms [see Fig. 1(c)] in the heptagon. Here, the double-bond length $d_{Sb_4-Sb_{1'}}$ = 2.712 Å is shorter than the single-bond lengths such as $d_{Sb_1-Sb_2}$ = 2.886 Å, $d_{Sb_2-Sb_3}$ = 2.914 Å, and $d_{Sb_3-Sb_4}$ = 2.865 Å (see Table I). This saturation of DBs significantly alters the band dispersion of edge states with a large gap opening of ∼1 eV [see Fig. 2(c)]. By using the nudged elastic-band method, we calculate the energy profile along the transition path from ZSNR-BE to ZSNR-PE and ZSNR-RE. The result is displayed in Fig. 4(a). We find that the total energy decreases monotonically on going from ZSNR-BE to ZSNR-PE, whereas the transition pathway from ZSNR-PE to ZSNR-RE has an energy barrier of ∼27 meV per edge (estimated from the employed nanoribbon containing 30 Sb atoms within the 1×1 unit cell). Therefore, the former structural phase transition belongs to second order, while the latter one first order. Based on the calculated energy barrier between ZSNR-PE and ZSNR-RE, we can say that the edge reconstruction with bond breakage and new bond formation would occur at a temperature of ∼100 K, if an Arrhenius-type activation process is assumed with the usual attempt frequency of ∼$10^{13}$ Hz.

Next, we optimize the atomic structures of ZSNR-BE, ZSNR-PE, and ZSNR-RE using the PBE+SOC calculation. The inclusion of SOC hardly changes the bond lengths of edge atoms by less than 0.005 Å (see Table I). Figure 4(a) also shows the calculated total energies of ZSNR-PE and ZSNR-RE relative to ZSNR-BE. We find that ZSNR-PE and ZSNR-RE are energetically more favored over ZSNR-BE by 2.0 and 3.9 meV/atom, respectively. These PBE+SOC values are slightly reduced compared to the corresponding PBE ones (2.5 and 4.6 meV/atom). Note that the PBE+SOC energy barrier from ZSNR-PE to ZSNR-RE is also reduced to be 16 meV per edge [see Fig. 4(a)]. The lower panels of Figs. 2(a), 2(b), and 2(c) show the PBE+SOC band structures of ZSNR-BE, ZSNR-PE, and ZSNR-RE, respectively. The spin splitting appears over the Brillouin zone, as the SOC lifts the spin degeneracy by breaking inversion symmetry at each edge[18,35,36]. It is, however, noted that (i) such spin-split subbands are still degenerate because of the two edge states arising from the opposite edges and (ii) ZSNR-BE (ZSNR-PE) has a small gap opening of 21 (19) meV even at a time-reversal invariant momentum (TRIM) point of $\Gamma$, caused by the interedge interaction [see the inset of Fig. 2(a)]. Around the other TRIM point of X, the band dispersion represents a Rashba-type spin splitting[35]. We fit the $k$-dependent dispersion of the spin-split subbands [indicated by the arrows in Figs. 2(a), 2(b), and 2(c)] with the momentum offset $\Delta k_R$ and the Rashba energy $E_R$ by using the Rashba spin-splitting eigenvalues $\varepsilon_{\pm} = \frac{\hbar^2 k^2}{2m^*} \pm \alpha_R k$, where $m^*$ is the electron effective mass and $\alpha_R$ the Rashba parameter[37]. The evaluated values of ($\Delta k_R$, $E_R$, $\alpha_R$) are (0.22 Å$^{-1}$, 0.07 eV, 1.32 eVÅ), (0.03, 0.01, 1.79), and (0.06, 0.01, 0.48) for ZSNR-BE, ZSNR-PE, and ZSNR-RE, respectively.

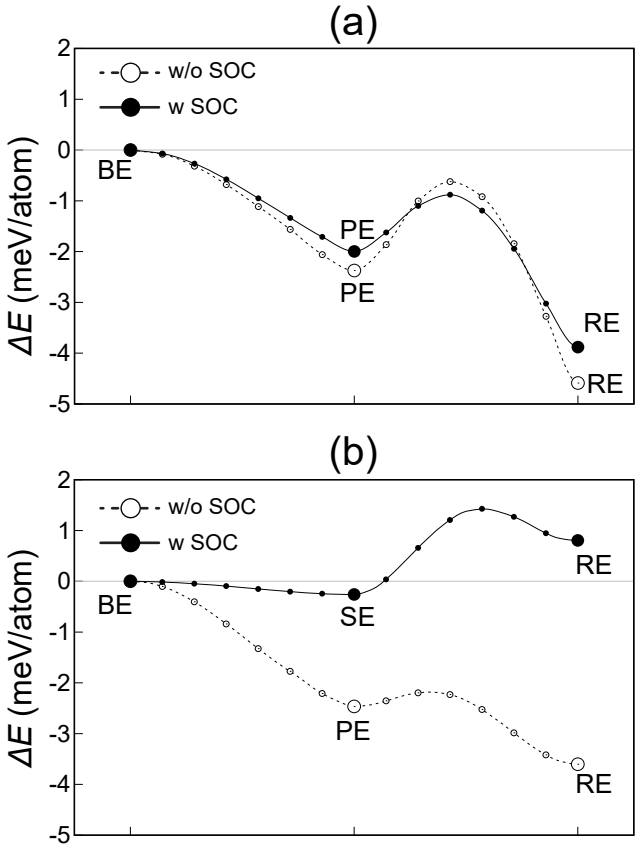

FIG. 4: (a) Total energies of ZSNR-PE and ZSNR-RE relative to ZSNR-BE, obtained without and with the inclusion of SOC. The corresponding ones in ZBNR are displayed in (b).

In order to obtain the spin texture of an isolated edge state, we consider the asymmetric nanoribbon structure by passivating one edge with H atoms. Figures 5(a) and 5(b) show the calculated band structures and spin textures of such H-passivated ZSNR-BE and ZSNR-PE, respectively. Unlike the above-mentioned symmetric nanoribbon cases, the spin-split edge state originating from the BE or PE side becomes nondegenerate everywhere except at the TRIM points of $\Gamma$ and X, where a hybridization gap opening due to interedge interaction does not occur. As shown in Fig. 5(a), the spin texture of edge state on the BE side has the $y$ and $z$ components of spin moments, which are perpendicular to the electron's wavevec-tor **k** pointing in the $x$ direction. Here, the spin texture satisfies the time-reversal symmetry that simultaneously reverses the wavevector and spin between the positive and negative $\Gamma$X lines. It is noteworthy that, due to the buckled geometry of ZSNR, the broken inversion symmetry at the BE side creates the potential gradients along the $y$ and $z$ directions, giving rise to an asymmetry of edge charge distribution. Since the edge states of ZSNR-BE are mostly composed of the Sb $p_y$ and $p_z$ orbitals [see Fig. 2(a)], such charge asymmetry is expected to produce the formation of orbital angular momentum (OAM) with the two components $L_y$ and $L_z$, which in turn determines the spin texture through the SOC[38]. Indeed, the calculated orbital magnetic moments in ZSNR-BE show the existence of two components $m_y$ and $m_z$ [see Fig. 5(c)]. Meanwhile, ZSNR-PE has all three components $m_x$, $m_y$, and $m_z$ for the spin and orbital magnetic moments [Figs. 5(b) and 5(d)]. Here, the presence of $m_x$ may be caused by breaking mirror-plane symmetry at the PE side. It is thus likely that the Rashba-type spin splitting of edge state can be associated with its unquenched OAM that is generated by a charge asymmetry at the edge[38].

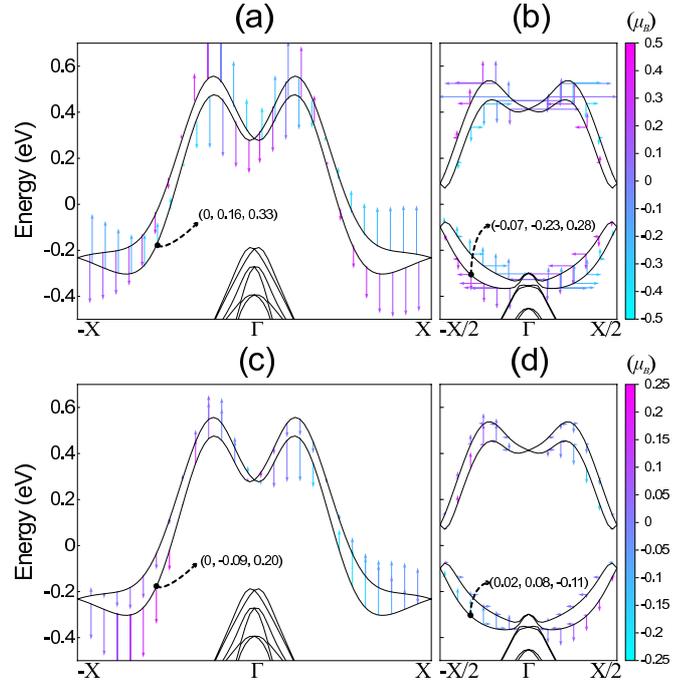

FIG. 5: (Color online) Spin textures of edge states in (a) ZSNR-BE and (b) ZSNR-PE, obtained using nanoribbon with the passivation of H atoms on one edge. The corresponding orbital magnetic moments for ZSNR-BE and ZSNR-PE are also displayed in (c) and (d), respectively. The horizontal and vertical arrows represent the components $m_x$ and $m_y$, respectively. Here, the vertical arrow representing the spin-moment component $m_y$ is reduced to one third of its original length. The colors of the arrows indicate $m_z$. The numbers represent the values (in $\mu_B$) of ($m_x$, $m_y$, $m_z$) at several **k** points.

### 2. Zigzag Bi nanoribbon

Similar to the case of ZSNR, we begin with the optimization of the BE, PE, and RE structures of ZBNR (termed

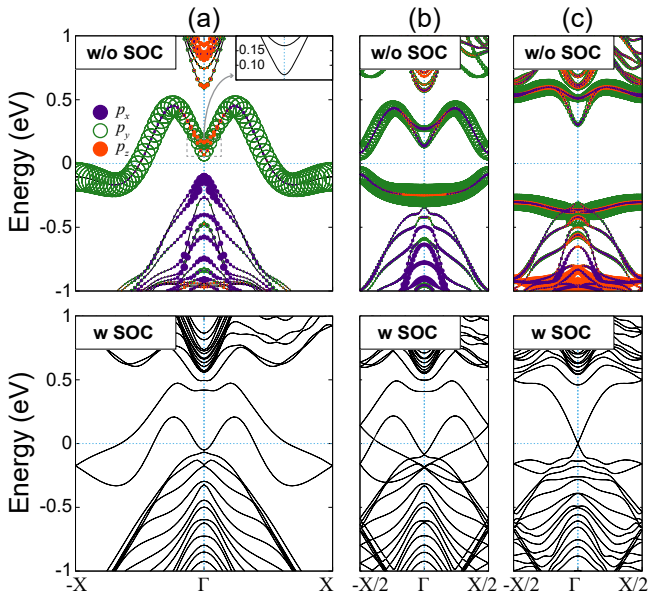

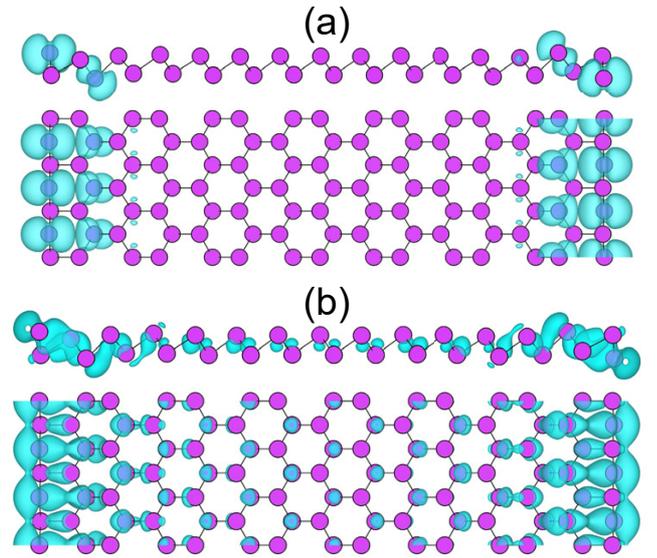

FIG. 6: (Color online) Band structures of (a) ZBNR-BE, (b) ZBNR-PE, and (c) ZBNR-RE, obtained without and with the inclusion of SOC. The inset in (a) magnifies the band dispersions near the Γ point. The bands projected onto the $6p_x$, $6p_y$ and $6p_z$ orbitals of the edge $Bi_1$ and $Bi_2$ atoms are displayed with circles whose radii are proportional to the weights of such orbitals.

FIG. 7: (Color online) Side and top views of the charge densities of the ZBNR-BE edge state at the (a) X and (b) Γ points. The charge densities are drawn with an isosurface of $5 \times 10^{-4}$ electrons/Å.

ZBNR-BE, ZBNR-PE, and ZBNR-RE, respectively) using the PBE calculation without the inclusion of SOC. The calculated bond lengths of edge atoms are listed in Table I. We find that ZBNR-PE and ZBNR-RE become more stable than ZBNR-BE by 2.4 and 3.6 meV/atom, respectively [see Fig. 4(b)]. This stabilization energy of ZBNR-PE is similar to that (2.5 meV) of ZSNR-PE, but the stabilization energy of ZBNR-RE and the energy barrier (∼5 meV per edge) from ZBNR-PE to ZBNR-RE are smaller than the corresponding ones (4.6 meV/atom and ∼27 meV per edge) in ZSNR [see Figs. 4(a) and 4(b)]. The calculated band structures of ZBNR-BE, ZBNR-PE, and ZBNR-RE are given in the upper panels of Figs. 6(a), 6(b), and 6(c), respectively. It is seen that ZBNR-BE has the half-filled bands with a Fermi surface nesting vector of $2k_F = \pi/a_x$, while ZBNR-PE (ZBNR-RE) is insulating with a gap opening of 169 (608) meV. Thus, we can say that ZBNR-PE and ZBNR-RE are stabilized by the band-gap openings due to the Fermi surface nesting-induced CDW formation and the pentagon-heptagon edge reconstruction, respectively. For ZBNR-BE, the charge character of edge state at the X point represents the localization of DB electrons near the edges [see Fig. 7(a)], while that at the Γ point shows a significantly larger penetration into the center of nanoribbon compared to ZSNR-BE [see Figs. 3(b) and 7(b)]. The resulting interedge interaction in ZBNR-BE gives rise to a hybridization gap of 99 meV around the Γ point [see the inset of Fig. 6(a)], which is larger than the corresponding one (14 meV) in ZSNR-BE.

Next, we perform PBE+SOC calculations to optimize the atomic structures of ZBNR-BE, ZBNR-PE, and ZBNR-RE.

The inclusion of SOC gives some small changes in the bond lengths of edge atoms (see Table I), but drastically influences the band structures of edge states [see the lower panels of Figs. 6(a), 6(b), and 6(c)] because the Bi(111) bilayer is transformed into a topological insulator[19,21,23,25]. We find that the edge states are characterized by an odd number of crossing across $E_F$ along the ΓX line (three for ZBNR-BE and ZBNR-PE, while one for ZBNR-RE), representing their non-trivial topological features[15,39]. By contrast, the edge states in ZSNR have even or zero crossing across $E_F$, i.e., two for ZSNR-BE and zero for ZSNR-PE and ZSNR-RE [see the lower panels of Figs. 2(a), 2(b), and 2(c)]. These different features of edge states between ZBNR and ZSNR reflect their contrasting bulk-edge correspondence of topological and normal insulators, respectively. For ZBNR-BE, we find a relatively larger Rashba-type spin splitting with $\Delta k_R = 0.22$ Å$^{-1}$ and $E_R = 0.16$ eV around the X point, compared to ZSNR-BE [see Figs. 2(a) and 6(a)]. Using the Rashba spin-splitting eigenvalues[37], the characteristic parameters of $\Delta k_R$ and $E_R$ in ZBNR-BE are fitted to evaluate $\alpha_R = 2.89$ eV Å, larger than the corresponding one (1.32 eV Å) in ZSNR-BE. Indeed, an angle-resolved photoemission spectroscopy experiment for Bi thin films on the Si(111) surface observed a giant Rashba spin-splitting[26]. Interestingly, it is noticeable that a $2k_F$ periodic lattice distortion disappears in ZBNR-PE, leading to the 1×1 shear-distorted edge structure (termed ZBNR-SE) with $d_{Bi_1-Bi_2} = d_{Bi_3-Bi_4} = 3.049$ Å and $d_{Bi_2-Bi_3} = d_{Bi_4-Bi_{1'}} = 3.020$ Å (see Table I). Considering the fact that the Peierls distortion is accompanied with a gap opening, the absence of a 2×1 lattice distortion in ZBNR-SE is likely to be caused by topologically protected gapless edge states. For ZBNR-RE, the dispersion of edge states is found to be much reshaped by reconstruction, i.e., the edge bands are not only moved out of the band-gap regime around the X/2 point but also cross $E_F$ at the Γ point [see

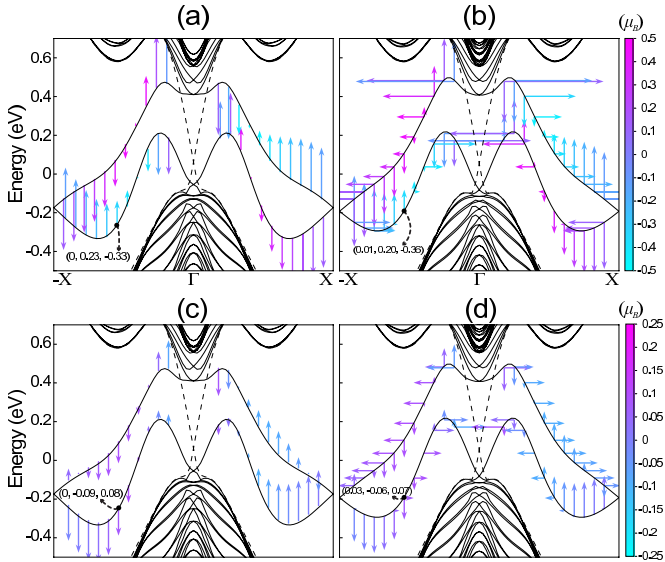

FIG. 8: (Color online) Spin textures of edge states in (a) ZBNR-BE and (b) ZBNR-PE, obtained using nanoribbon with the passivation of H atoms on one edge. The corresponding orbital magnetic moments for ZSNR-BE and ZSNR-PE are also displayed in (c) and (d), respectively. The results for ZBNR-SE are plotted within the Brillouin zone of the 1×1 unit cell. The dashed lines represent the edge bands originating from the H-passivated side. The horizontal and vertical arrows represent the components $m_x$ and $m_y$, respectively. Here, the vertical arrow representing the spin-moment component $m_y$ is reduced to one third of its original length. The colors of the arrows indicate $m_z$. The numbers represent the values (in $\mu_B$) of $(m_x, m_y, m_z)$ at several **k** points.

Fig. 6(c)]. Thus, we can say that, although the edge states of ZBNR are robust in their topological nature (e.g., no backscattering on conductivity), edge reconstruction not only significantly changes the dispersion shape of edge states but also manipulates the number of edge conduction channels[23,25].

Figure 4(b) also shows the PBE+SOC total energies of ZBNR-SE and ZBNR-RE relative to ZBNR-BE. We find a difference in their stabilities compared to the PBE results: i.e., ZBNR-SE is slightly more stable than ZSNR-BE by 0.3 meV/atom, while ZBNR-RE is less stable than ZBNR-BE by 0.8 meV/atom. Therefore, ZBNR-SE becomes the most stable structure, indicating a significant SOC effect on the stability of competing edge structures. This effect can be attributed to the SOC-induced quantum phase transition of Bi(111) bilayer from a normal to a topological insulator. Specifically, ZBNR-RE has the largest band gap in the normal insulating state [see the upper panel of Fig. 6(c)], but, when it becomes a topological insulator by SOC, the presence of topologically protected gapless edge states loses more electronic energy than the cases of ZBNR-BE and ZBNR-SE. This SOC-induced switching of the ground structure in ZBNR is distinct from ZSNR, where the inclusion of SOC does not change the stability of edge structures [see Fig. 4(a)].

Finally, we obtain the spin texture of an isolated edge state using the asymmetric nanoribbon structure with the passivation of H atoms on one edge. Figures 8(a) and 8(b) show the calculated band structures and spin textures of ZBNR-BE and ZBNR-SE, respectively. Here, the unfolded band structure of ZBNR-SE is plotted within the Brillouin zone of the 1×1 unit cell. It is noted that the edge bands originating from the H-passivated side [represented by the dashed lines in Figs. 8(a) and 8(b)] are well separated from those from the BE and SE sides[25]. We find that the spin texture of one edge state in ZBNR-BE (having the mirror-plane symmetry with the $yz$ plane) has the two components $m_y$ and $m_z$ [see Fig. 8(a)], whereas that in ZBNR-SE (breaking the mirror-plane symmetry) involves all three components $m_x$, $m_y$, and $m_z$ [Fig. 8(b)]. Further, the calculated orbital magnetic moments also show the presence of $m_y$ and $m_z$ in ZBNR-BE [see Fig. 8(c)], while $m_x$, $m_y$, and $m_z$ in ZBNR-SE [Fig. 8(d)]. Therefore, the edge states of ZBNR-BE (ZBNR-SE) have the spin texture and orbital magnetic moments that are analogous to those of ZSNR-BE (ZSNR-PE).

## IV. SUMMARY

We have investigated competing edge structures of zigzag Sb(111) and Bi(111) nanoribbons using the DFT calculations with/without the inclusion of SOC. We found that the edge structures of the two nanoribbons are drastically changed with respect to their topologically trivial and nontrivial edge states. For zigzag Sb(111) nanoribbon, SOC does not affect the relative stability of several edge structures: i.e., the Peierls-distorted or reconstructed edge structure with a bandgap opening is more stable than the metallic bulk-truncated edge structure. However, for zigzag Bi(111) nanoribbon, SOC drastically changes the stability order of such edge structures: i.e., the absence of SOC gives the same order as predicted in zigzag Sb(111) nanoribbon, while the inclusion of SOC favors the metallic shear-distorted edge structure rather than the Peierls-distorted and reconstructed edge structures. The present findings are rather generic and hence, they should be more broadly applicable to determine the edge structures of other normal and topological insulators.

**Acknowledgement.** This work was supported by National Research Foundation of Korea (NRF) grant funded by the Korean Government (Grant Nos. 2015M3D1A1070639 and 2016K1A4A3914691). The calculations were performed by KISTI supercomputing center through the strategic support program (KSC-2017-C3-0041) for the supercomputing application research.S.-M. J., S. Y., and H.-J. K. contributed equally to this work.
* Corresponding author: chojh@hanyang.ac.kr